\documentclass[fleqn,usenatbib]{mnras}

\usepackage{newtxtext,newtxmath}

\usepackage[T1]{fontenc}
\usepackage{ae,aecompl}

\usepackage{graphicx}	
\usepackage{amsmath}	
\usepackage{amssymb}	
\usepackage{paralist}
\usepackage{blindtext}

\title[An arc-length approximation for elliptical orbits]{AN ARC-LENGTH APPROXIMATION FOR ELLIPTICAL ORBITS}

\author[A. B. Karki et al.]{
Ashim B. Karki$^{1}$\thanks{E-mail: ashimkarki54@gmail.com}
and Aayush Jha$^{2}$\thanks{E-mail: 32577aayushjha@gmail.com}
\\
$^{1}$Wolfram Research, Champaign, IL, USA\\
$^{2}$Department of Physics, St. Xavier's College, Maitighar, Kathmandu 44600, Nepal\\
}

\date{Accepted XXX. Received YYY; in original form ZZZ}

\pubyear{2019}

\begin{document}
\label{firstpage}
\pagerange{\pageref{firstpage}--\pageref{lastpage}}
\maketitle

\begin{abstract}
In this paper, we overlay a continuum of analytical relations which essentially serve to compute the arc-length described by a celestial body in an elliptic orbit within a stipulated time interval. The formalism is based upon a two-dimensional heliocentric coordinate frame, where both the coordinates are parameterized as two infinitely differentiable functions in time by using the Lagrange inversion theorem. The parameterization is firstly endorsed to generate a dynamically consistent ephemerides for any celestial object in an elliptic orbit, and thereafter manifested into a numerical integration routine to approximate the arc-lengths delineated within an arbitrary interval of time. As elucidated, the presented formalism can also be orchestrated to quantify the perimeters of elliptic orbits of celestial bodies solely based upon their orbital period and other intrinsic characteristics.
\end{abstract}




\section{Introduction} \label{introduction}
Over the past few decades, the advancement in astrodynamics has instigated a new era of celestial ephemerides. With the advent of novel observational routines - namely, Microwave ranging, lunar laser ranging, VLBI measurements, etc. - astronomers have not only subsided ambiguity in estimations of orbital elements of celestial objects but have also deduced results that are consequently more viable and pragmatic. Throughout the course of its refinement, data procured from theoretical approaches for determination of ephemerides has advanced to a machine-level precision with the incorporation of modern computational software. The refurbishment of classical analytical theories proposed by Newcomb \citep{newcomb1882discussion}, rendered to make reinvestigations into the Nautical Almanac, eventually came to a halt in the late 1960s. Inevitably, digital computation was set on par with modern astronomy when \cite{eckert1951coordinates} pioneered to numerically integrate the equations of orbital motion of the planets and the Moon, thereafter followed by \cite{oesterwinter1972new}. The enhancement course was overtaken by American groups at JPL, Pasadena \citep{standish1976jpl, newhall1983102} with the commencement of a series of Development Ephemerides (which took into account empirical discrepancies, such as perturbations, obliquity, nutations, and librations of the planets) as the need for more accurate planetary ephemerides became evident with the advancement in space exploration, and a plethora of other rectifications have been made in this arena hitherto.\\
Notwithstanding the meticulously stringent observations made till date, the contemporary ephemerides are mostly modelled upon the Earth-centred inertial (commonly, the J2000) coordinate frames for numerical integration, whereas, a miniscule progression has been made to ameliorate corresponding concepts reliant upon heliocentric coordinate system. In this paper, we set forth analytical relations to compute the positions, and ultimately the arc-lengths of celestial bodies by initially parameterizing the elliptic (in this case, heliocentric) coordinates as direct functions of time elapsed from the epoch. The parameterization that follows can be thought of as a rudimentary application of the open-form solution to the Inverse Kepler\textquotesingle s problem as it makes a predominant use of the Lagrange Inversion theorem to revert the Kepler\textquotesingle s equation and express the eccentric anomaly in terms of mean anomaly and hence, secure two parametric functions in time. The first order time-derivatives of these functions, in the light of the inverted series of the eccentric anomaly, can thereafter communally yield an expression for arc-length traced by the celestial body in between two arbitrary time periods referenced from the epoch. As a corroboration to our work, values of the orbital elements of Earth-Moon barycentre and other celestial bodies - rectified to an eighth decimal-digit precision - have been deployed from \cite{murray1999solar} to numerically integrate the afore-mentioned expression, and hence compute the perimeter of their elliptic orbits in the form of an arc-length approximation.\\
In Section \ref{dis}, the Kepler\textquotesingle s equation is inverted using the Lagrange inversion theorem, supplemented with functions of eccentricity anomaly and its first order time-derivative. In Section \ref{procession}, the heliocentric coordinates are parametrized in time and ephemerides of Earth-Moon barycentre is tabulated along with an illustration. Section \ref{arclength} constructs a general expression to numerically compute the arc-lengths described by celestial bodies advancing in elliptic orbits. In Section \ref{rendition}, we demonstrate how this expression can be effectuated to calculate the perimeters of elliptic orbits. Section \ref{apsidal} scrutinizes about the probable causes of ambiguities in approximations secured by the equations deduced in this paper. The final verdicts are demarcated in the last section. 

\section{DISAMBIGUATION OF THE INVERSE KEPLER\textquotesingle S EQUATION}\label{dis}
On account of the fact that the Kepler\textquotesingle s equation cannot be analytically solved for eccentric anomaly \textit{E}, a myriad of fixed-point iterative solutions, which employ the Newton-Raphson method, have been proposed and revised so far. However, it is worthwhile to note that the Lagrange inversion theorem is undeniably a more pertinent approach in this case, as it facilitates the reversion of the equation by expressing it into an open-form convergent Maclaurin series in terms of the mean anomaly \textit{M}, allowing it to be solved a priori. The accuracy of the series can thereby be optimized by truncating it as per convenience. The relation between mean anomaly \textit{M} and the eccentric anomaly \textit{E} can be set up in accordance to the Kepler\textquotesingle s equation as
\begin{equation}\label{eq1}
M=h(E)=E-e \sin E,
\end{equation}
where \textit{e} is the eccentricity of the elliptic orbit and \textit{h(E)} represents a function of\textit{ M} in terms of $\textit{E}$. The function is clearly analytic and differentiable at$\textit{ E=0}$, and as $\textit{h}^{\prime}(0)=1-e \neq 0, \textit{h}(0)=0$, \textit{E }can reasonably be enumerated in the form of \textit{M} as
\begin{equation}\label{eq2}
E=\sum_{k=1}^{\infty} \frac{M^{k}}{k !} \lim _{E \rightarrow 0^{+}}\left(\frac{\mathrm{d}^{k-1}}{\mathrm{d} E^{k-1}}\left(\left(\frac{E}{E-e \sin E}\right)^{k}\right)\right), \forall e \neq 1.
\end{equation}
On evaluation, the equation (\ref{eq2}) yields
\begin{multline}\label{eq3}
E=\frac{M}{1-e}-\frac{e}{(1-e)^{4}} \frac{M^{3}}{3 !}+\frac{\left(9 e^{2}+e\right)}{(1-e)^{7}} \frac{M^{5}}{5 !}\\
-\frac{\left(225 e^{3}+54 e^{2}+e\right)}{(1-e)^{10}} \frac{M^{7}}{7 !}+\mathcal{O}\left(M^{9}\right), \forall e \neq 1.
\end{multline}
The mean anomaly $M$ can also be rendered as $M=nt$, where \textit{t} is the time elapsed since pericentre passage and the mean motion\footnote{The standard gravitational parameter is denoted here by $\mu=\mathcal{G}\left(m_{1}+m_{2}\right)$, where $m_{1}$ and $m_{2}$ are the two corresponding masses of the bodies in the orbiting system and $\mathcal{G}$ is the Newtonian Gravitational constant. The orbital period of the celestial body is denoted by\textit{ P}, \textit{a} is its semi-major axis. }$n=\sqrt{\mu / a^{3}}=2 \pi / P.$\\
The eccentricity anomaly \textit{E} can now be deduced in terms of \textit{t} by reformulating equation (\ref{eq3}) as
\begin{multline}\label{eq4}
E=\frac{n t}{1-e}-\frac{e}{(1-e)^{4}} \frac{n^{3} t^{3}}{3 !}+\frac{\left(9 e^{2}+e\right)}{(1-e)^{7}} \frac{n^{5} t^{5}}{5 !}\\
-\frac{\left(225 e^{3}+54 e^{2}+e\right)}{(1-e)^{10}} \frac{n^{7} t^{7}}{7 !}+\mathcal{O}\left(t^{9}\right).
\end{multline}
Taking the first order time-derivative of equation (\ref{eq4}) on a term-by-term basis leads to
\begin{multline}\label{eq5}
\dot{E}=\frac{n}{1-e}-\frac{e}{(1-e)^{4}} \frac{n^{3} t^{2}}{2 !}+\frac{\left(9 e^{2}+e\right)}{(1-e)^{7}} \frac{n^{5} t^{4}}{4 !}\\
-\frac{\left(225 e^{3}+54 e^{2}+e\right)}{(1-e)^{10}} \frac{n^{7} t^{6}}{6 !}+\mathcal{O}\left(t^{8}\right).
\end{multline}

\section{PROCESSION OF EARTH IN RESPECT OF HELIOCENTRIC COORDINATE FRAMES}\label{procession}

Affixing the Sun at one of the two foci of the elliptic orbit, it can be derived from general geometry that the heliocentric cartesian coordinates\footnotemark  of the Earth-Moon barycentre during the procession are
\footnotetext{As a matter of convenience, the orbital inclination can be set to 0 for computation and the ascending node can be placed in the reference direction allowing the coordinate frames to be parsed in a cartesian plane.}
\begin{equation}\label{eq6}
x=a(\cos E-e),
\end{equation} 
\begin{equation}\label{eq7}
y=b \sin E,
\end{equation}
where \textit{a} is the semi-major axis of the elliptical orbit and \textit{b} is its semi-minor axis. Substituting the expression for \textit{E} in equations (\ref{eq6}) - (\ref{eq7}), we get
\begin{equation}\label{eq8}
x=a \cos \left[\sum_{k=1}^{\infty} \frac{M^{k}}{k !} \lim _{E \rightarrow 0^{+}}\left(\frac{\mathrm{d}^{k-1}}{\mathrm{d} E^{k-1}}\left(\left(\frac{E}{E-e \sin E}\right)^{k}\right)\right)\right]-a e,
\end{equation}
\begin{equation}\label{eq9}
y=b \sin \left[\sum_{k=1}^{\infty} \frac{M^{k}}{k !} \lim _{E \rightarrow 0^{+}}\left(\frac{\mathrm{d}^{k-1}}{\mathrm{d} E^{k-1}}\left(\left(\frac{E}{E-e \sin E}\right)^{k}\right)\right)\right],
\end{equation}
which on further simplification can be written respectively as
\begin{multline}\label{10}
    x=a \cos \Bigg[\frac{M}{1-e}-\frac{e}{(1-e)^{4}} \frac{M^{3}}{3 !}+\frac{\left(9 e^{2}+e\right)}{(1-e)^{7}} \frac{M^{5}}{5 !}\\
    -\frac{\left(225 e^{3}+54 e^{2}+e\right)}{(1-e)^{10}} \frac{M^{7}}{7 !}+\mathcal{O}\left(M^{9}\right)\Bigg]-a e,
\end{multline}

\begin{multline}\label{11}
y=b \sin \Bigg[\frac{M}{1-e}-\frac{e}{(1-e)^{4}} \frac{M^{3}}{3 !}+\frac{\left(9 e^{2}+e\right)}{(1-e)^{7}} \frac{M^{5}}{5 !}\\
-\frac{\left(225 e^{3}+54 e^{2}+e\right)}{(1-e)^{10}} \frac{M^{7}}{7 !}+\mathcal{O}\left(M^{9}\right)\Bigg].
\end{multline}

In order to manifest the elliptical orbit as a parametric curve such that the set of arbitrary points $(x, y)=(f(t), g(t))$ describe a continuous and differentiable trajectory, $0 \leq t \leq P$, one can plug in the series of $E$ secured from equation (\ref{eq4}) into equations (\ref{eq6}) - (\ref{eq7}) to write\footnotemark
\footnotetext{It is to be noted that, from the manner in which the equations have been derived, both $E$ and $M$ should be expressed in circular measure.}
\begin{multline}\label{eq12}
x=f(t)=a \cos \Bigg[\frac{n t}{1-e}-\frac{e}{(1-e)^{4}} \frac{n^{3} t^{3}}{3 !}+\frac{\left(9 e^{2}+e\right)}{(1-e)^{7}} \frac{n^{5} t^{5}}{5 !}\\
-\frac{\left(225 e^{3}+54 e^{2}+e\right)}{(1-e)^{10}} \frac{n^{7} t^{7}}{7 !}+O\left(t^{9}\right)\Bigg]-a e,
\end{multline}
\begin{multline}\label{eq13}
y=g(t)=b \sin \Bigg[\frac{n t}{1-e}-\frac{e}{(1-e)^{4}} \frac{n^{3} t^{3}}{3 !}+\frac{\left(9 e^{2}+e\right)}{(1-e)^{7}} \frac{n^{5} t^{5}}{5 !}\\
-\frac{\left(225 e^{3}+54 e^{2}+e\right)}{(1-e)^{10}} \frac{n^{7} t^{7}}{7 !}+\mathcal{O}\left(t^{9}\right)\Bigg].
\end{multline}

\begin{table}
	\centering
	
	\caption{If one centres the origin (here, the centre of mass of the Sun) at the focus of the elliptical orbit and overlooks perturbative effects of other planets and apsidal procession of the orbit, then the course of maneuver of the Earth-Moon barycentre can be described using equations (\ref{eq12}) - (\ref{eq13}). The variations in the heliocentric coordinates are quantified below apropos of their corresponding arbitrary time periods $t$ (time elapsed since periastron passage). 
	Note that $t$ has been expressed in Earth days (d), and the horizontal and vertical displacements from the origin have been computed in Astronomical Units (AU).}
		\begin{tabular}{rrr} 
		\hline
		\hline
		$t$ & $x$ & $y$ \\
		\hline
		30 & 0.849832888 & 0.499037433 \\
		60 & 0.485074866 & 0.864872995 \\
		90 & -0.013617045 & 0.999852941 \\
		120 &-0.513134358 & 0.867961230 \\
		150 & -0.880147059 & 0.504387701 \\
		180 & -1.016691176 & 0.006185495 \\
		210 & -0.886323529 &-0.493667781\\
		240 & -0.523836898 &-0.861751337 \\
		270 & -0.025989706 &-0.999819519 \\
		300 & 0.474334225 &-0.871016043 \\
		330 & 0.843475936 &-0.509718583\\
		360 & 0.983215241 & -0.012370722 \\
		\hline
		\hline
	\end{tabular}
	\label{table1}
\end{table}

\begin{figure}
	
	\includegraphics[width=\columnwidth]{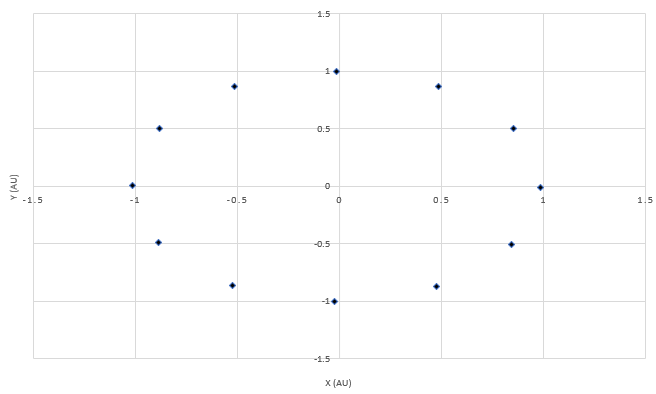}
    \caption{The scatter represents the positions of the Earth-Moon barycentre at varying time periods $t$. At this point, it can be inferred that the Earth-Moon barycentre reaches back to the periastron after 360 Earth days (approximately, one year) have passed.}
    \label{figure1}
\end{figure}
\section{ARC-LENGTHS OF ELLIPTICAL ORBITS: NUMERICAL INTEGRATION ROUTINE}\label{arclength}
Contingent upon the relations we have established so far, here we construct a generalized equation that generates the arc-length covered by a celestial object in between any two arbitrary times - both referenced from the periastron passage. The first order time-derivatives of the heliocentric coordinates can be maintained from equations (\ref{eq6}) - (\ref{eq7}) as
\begin{equation}\label{eq14}
    \dot{x}=-a \cdot \dot{E} \sin E,
\end{equation}
\begin{equation}\label{eq15}
    \dot{y}=b \cdot \dot{E} \cos E.
\end{equation}
One can now presume the elliptical orbit to be a parametrically defined curve continuously differentiable on the interval $0 \leq t \leq P$ in consonance with equations (\ref{eq12}) - (\ref{eq13}). Also, $f^{\prime}(t)$ and $g^{\prime}(t)$ are not simultaneously zero, as every ellipse is devoid of cusps or corners. In account of its geometry, the elliptical orbit is transversed exactly once and does not double back on itself or reverse its direction of motion, since $\left(f^{\prime}\right)^{2}+\left(g^{\prime}\right)^{2}>0$ throughout the interval $0 \leq t \leq P$. Therefore, one can satisfactorily define the arc-length in this case to be
\begin{equation}\label{eq16}
    \ell=\int_{t_{1}}^{t_{2}} \sqrt{\dot{x}^{2}+\dot{y}^{2}} {\mathrm{d}t},
\end{equation}
where $\ell$ represents the arc-length covered by the celestial object within two arbitrary intervals of time $t_{1}$ and $t_{2}$ commencing from the periastron passage. From equations (\ref{eq14}) - (\ref{eq15}) one can modify equation (\ref{eq16}) into
\begin{equation}\label{eq17}
    \ell=\int_{t_{1}}^{t_{2}} \dot{E} \sqrt{a^{2} \sin ^{2} E+b^{2} \cos ^{2} E} {\mathrm{d}t}.
\end{equation}
From general geometry, the relationship between semi-major axis $a$, semi-minor axis $b$, and the eccentricity $e$ of an ellipse is
\begin{equation}\label{eq18}
    e=\sqrt{1-\frac{b^{2}}{a^{2}}}.
\end{equation}
One can modulate equation (\ref{eq17}) purely in terms of eccentricity $e$ and semi-major axis $a$ using equation (\ref{eq18}) as
\begin{equation}\label{eq19}
    \ell=a \int_{t_{1}}^{t_{2}} \dot{E} \sqrt{1-e^{2} \cos ^{2} E}{\mathrm{d}t}.
\end{equation}
Using equation (\ref{eq2}), one can write equation (\ref{eq19}) as 
\begin{multline}\label{eq20}
    \ell=a \int_{t_{1}}^{t_{2}} \frac{\mathrm{d}}{\mathrm{d} t}\left(\sum_{k=1}^{\infty} \frac{M^{k}}{k !} \lim _{E \rightarrow 0^{+}}\left(\frac{\mathrm{d}^{k-1}}{\mathrm{d} E^{k-1}}\left(\left(\frac{E}{E-e \sin E}\right)^{k}\right)\right)\right)\cdot\\
    \sqrt{1-e^{2} \cos ^{2}\left[\sum_{k=1}^{\infty} \frac{M^{k}}{k !} \lim _{E \rightarrow 0^{+}}\left(\frac{\mathrm{d}^{k-1}}{\mathrm{d} E^{k-1}}\left(\left(\frac{E}{E-e \sin E}\right)^{k}\right)\right)\right]}{\mathrm{d}t} .
\end{multline}
On expansion of equation (\ref{eq20}) and successive substitution of definition of $M=n t$ (see also equations \ref{eq4} and \ref{eq5}), one can get
\begin{multline}\label{eq21}
    \ell=a \int_{t_{1}}^{t_{2}}\bigg(\frac{n}{1-e}-\frac{e}{(1-e)^{4}} \frac{n^{3} t^{2}}{2 !}+\mathcal{O}\left(t^{4}\right)\bigg)\cdot\\
    \sqrt{1-e^{2} \cos ^{2}\left[\frac{n t}{1-e}-\frac{e}{(1-e)^{4}} \frac{n^{3} t^{3}}{3 !}+\mathcal{O}\left(t^{5}\right)\right]} {\mathrm{d}t}. 
\end{multline}
Hereafter, one can numerically integrate equation (\ref{eq21}) to estimate the length of the arc advanced by any celestial object in its elliptical orbit. In order to comply with the Second law of Kepler, the equations have been preliminarily framed in a way that numerically integrated arc-lengths are relatively higher when the celestial body is closer to the periapsis than when it is approaching the apoapsis for equal durations of time; the effect is ensued as a result of tangential velocity variations throughout the course of its orbit. (see also table \ref{table2}). 

\begin{table}
	\centering
	
	\caption{The arc-lengths delineated by the Earth-Moon barycentre at discrete intervals of time $\Delta t$ for approximately one half of the elliptic orbit (containing the apoapsis) are demonstrated. Within these eighteen different time intervals (instigating from the 90th day to the 270th day), one can notice that, even though the time intervals are evenly spaced, the arc-lengths first gradually diminish until the Earth-Moon barycentre has reached the apoapsis (around 180th day), and then moderately increase as the intervals proceed.}
	\begin{tabular}{cc} 
		\hline
		\hline
		Time-interval $\Delta t$ (Earth days) & Arc-length (AU) \\
		\hline
		90 to 100 & 0.173978721  \\
		100 to 110 & 0.173977334  \\
		110 to 120 & 0.173974685 \\
		120 to 130 & 0.173971091  \\
		130 to 140 & 0.173966982  \\
		140 to 150  & 0.173963285  \\
		150 to 160  & 0.173959195 \\
		160 to 170  & 0.173956450  \\
		170 to 180 & 0.173954945  \\
		180 to 190 & 0.173954862  \\
		190 to 200 & 0.173956209 \\
		200 to 210  & 0.173958827  \\
		210 to 220 & 0.173962399 \\
		220 to 230 & 0.173966500 \\
		230 to 240 & 0.173970637 \\
		240 to 250 & 0.173974314 \\
		250 to 260 & 0.173977090 \\
		260 to 270 & 0.173978633 \\
		\hline
		\hline
	\end{tabular}
	\label{table2}
\end{table}

\section{RENDITION OF THE CIRCUMFERENCE OF THE ELLIPTIC ORBIT AS AN ARC-LENGTH}\label{rendition}
The parametrically defined arc-length obtained from equation (\ref{eq21}) must result in the total circumference of the elliptical orbit when numerically integrated within the limits 0 to $P$, where $P$ is the orbital period of the celestial body. If the origin is set at the elliptic centre, one can parametrically define the elliptic orbit in polar form to be
\begin{equation}\label{eq22}
  u=a \sin \theta, 
\end{equation}
\begin{equation}\label{eq23}
    v=b \cos \theta,
\end{equation}
where $a$ is its semi-major axis, and $b$ is its semi-minor axis such that $a>b$ and $0 \leq \theta \leq 2 \pi$. Then, 
\begin{equation}\label{eq24}
    \left(\frac{\mathrm{d} u}{\mathrm{d} \theta}\right)^{2}+\left(\frac{\mathrm{d} v}{\mathrm{d} \theta}\right)^{2}=a^{2} \cos ^{2} \theta+b^{2} \sin ^{2} \theta,
\end{equation}
which from equation (\ref{eq18}) can be written as
\begin{equation}\label{eq25}
    \left(\frac{\mathrm{d} u}{\mathrm{d} \theta}\right)^{2}+\left(\frac{\mathrm{d} v}{\mathrm{d} \theta}\right)^{2}=a^{2}\left(1-e^{2} \sin ^{2} \theta\right),
\end{equation}
where $e$ is the orbit\textquotesingle s eccentricity. Then, the complete elliptic integral of second kind (perimeter of the ellipse), $\wp$, can be given as
\begin{equation}\label{eq26}
    \wp=4 a \int_{0}^{\pi / 2} \sqrt{1-e^{2} \sin ^{2} \theta} {\mathrm{d}\theta}.
\end{equation}
The integral in equation (\ref{eq26}) can be evaluated after binomial expansion of $\sqrt{1-e^{2} \sin ^{2} \theta}$ to 
\begin{equation}\label{eq27}
    \wp=2 \pi a\left[1-\left(\frac{1}{2}\right)^{2} e^{2}-\left(\frac{1 \cdot 3}{2 \cdot 4}\right)^{2} \frac{e^{4}}{3}-\left(\frac{1 \cdot 3 \cdot 5}{2 \cdot 4 \cdot 6}\right)^{2} \frac{e^{6}}{5}-\mathcal{O}\left(e^{8}\right)\right]. 
\end{equation}
Finally, the celestial body will have covered an arc-length equal to the circumference of the elliptical orbit, so $\ell=\wp$, when numerically integrated within the limits 0 to $P$. From equations (\ref{eq21})-(\ref{eq27}), one can write
\begin{multline}\label{eq28}
    \int_{0}^{P}\bigg(\frac{n}{1-e}-\frac{e}{(1-e)^{4}} \frac{n^{3} t^{2}}{2 !}+\mathcal{O}\left(t^{4}\right)\bigg)\cdot\\
    \sqrt{1-e^{2} \cos ^{2}\left[\frac{n t}{1-e}-\frac{e}{(1-e)^{4}} \frac{n^{3} t^{3}}{3 !}+\mathcal{O}\left(t^{5}\right)\right]} {\mathrm{d}t}\\
    =2 \pi\left[1-\left(\frac{1}{2}\right)^{2} e^{2}-\left(\frac{1 \cdot 3}{2 \cdot 4}\right)^{2} \frac{e^{4}}{3}-\mathcal{O}\left(e^{6}\right)\right].
\end{multline}
\begin{table}
	\centering
	
	\caption{The circumferences of the elliptic orbits of five celestial bodies (with the least orbital inclinations) are computed and enlisted below alongside their deviations from standard measurements. }
	
	\begin{tabular}{ccc} 
		\hline
		\hline
		Celestial body & Circumference (AU) & Uncertainty (\%) \\
		\hline
		Earth & 6.354241 & 1.137\%  \\
		Neptune & 190.728037 & 0.955\%  \\
	    Adrastea (satellite)  & 0.005370 & 0.866\% \\
		Metis (satellite) & 0.005376 & 0.013\%  \\
		Atlas (satellite) & 0.005787 & 0.113\%  \\
	    \hline
		\hline
	\end{tabular}
	\label{table3}
\end{table}

\section{APSIDAL PROCESSION AND TRUNCATION ERRORS}\label{apsidal}
It is customary that the argument of periapsis in every elliptical orbit must experience a steady procession over time resulted by either general relativistic effects, rotational quadrupole bulges on the planet, stellar quadrupole moments, Lidov-Kozai mechanism, or perturbative effects of other planets \citep{kipping2011transits}. The work presented so far relies upon an isolated orbiting system; however, perturbations can ensue numerous, yet insignificant, discrepancies onto the existent conditions. \\
The inconsistencies induced in the computed estimates tend to vanish off when orbits with low inclinations are designated. The series can be truncated to any degree of accuracy, and the use of robust computational software should suffice to curtail any obscurity in calculations.

\section{Conclusions}

Parameterization of heliocentric coordinates as functions of angular quantities is unequivocally a straightforward task at hand, as they can be reformulated using elementary geometrical relations. However, this paper devises a new framework to revisit such conventionality by parameterizing the coordinates fundamentally in terms of time. This approach, in assent with the Lagrange inversion theorem, is then reinforced into a set of analytical relations predominantly governed by time. Complementary to the interposition of a new module for ephemerides calculations, this procedure also doubles as a numerical integration routine to compute arc-lengths traced by celestial bodies advancing in elliptical orbits.

\section*{Acknowledgements}

The author acknowledges the support conferred by Wolfram Research Inc., Champaign, Illinois, USA. Tabulated and graphical data presented throughout this paper has been exclusively computed using Mathematica.  






\bibliographystyle{mnras}
\bibliography{def}


\bsp	
\label{lastpage}
\end{document}